\documentclass[10pt,conference]{IEEEtran}
\IEEEoverridecommandlockouts
\usepackage{amsmath,amssymb,amsfonts}
\usepackage{graphicx}
\usepackage{textcomp}
\usepackage{xcolor}
\usepackage{tikz}
\usetikzlibrary{arrows.meta,positioning,fit,shapes.geometric}
\usepackage{hyperref}
\usepackage{xspace}

\newcommand{\secref}[1]{Section~\ref{#1}}
\newcommand{\figref}[1]{Figure~\ref{#1}}
\newcommand{\tabref}[1]{Table~\ref{#1}}
\newcommand{\lstref}[1]{Listing~\ref{#1}}

\newcommand{\ie}{\emph{i.e.},\xspace}

\newcommand{\etal}{\emph{et al.}\xspace}

\usepackage{listings}
\newcommand{\apn}{\textsc{SBVul}}

\usepackage[backend=biber,style=numeric,sorting=nyt,doi=true,url=false,eprint=false,isbn=false]{biblatex}

\def\BibTeX{{\rm B\kern-.05em{\sc i\kern-.025em b}\kern-.08em
    T\kern-.1667em\lower.7ex\hbox{E}\kern-.125emX}}

\begin{document}

\title{SBVUL: Statement-Based Vulnerability Detection \\
\thanks{
This research was funded in whole or in part by the Austrian Science Fund (FWF) 10.55776/P36698. For open access purposes,
the author has applied a CC BY public copyright license to any author accepted manuscript version arising from this submission.
The research was supported by the Austrian ministries BMIMI, BMWET and the State of Upper Austria in the frame of the SCCH COMET competence
center INTEGRATE (FFG 892418).}
}

%%\author{\IEEEauthorblockN{\textit{Anonymous Authors}}}

\author{\IEEEauthorblockN{Martin Pinzger, Katharina Stengg, Christian Macho}
\IEEEauthorblockA{
%\textit{Department of Informatics Systems} \\
\textit{University of Klagenfurt}\\
Klagenfurt, Austria \\
\{martin.pinzger, katharina.stengg, christian.macho\}@aau.at}
\and
\IEEEauthorblockN{Stefan Schöberl}
\IEEEauthorblockA{
%\textit{Department of Informatics Systems} \\
\textit{Software Competence Center Hagenber}\\
Klagenfurt, Austria\\
stefan.schoeberl@scch.at}
%\and
%\IEEEauthorblockN{Katharina Stengg}
%\IEEEauthorblockA{
%%\textit{Department of Informatics Systems} \\
%\textit{University of Klagenfurt}\\
%Klagenfurt, Austria\\
%katharina.stengg@aau.at}
%\and
%\IEEEauthorblockN{Christian Macho}
%\IEEEauthorblockA{
%%\textit{Department of Informatics Systems} \\
%\textit{University of Klagenfurt}\\
%Klagenfurt, Austria\\
%christian.macho@aau.at}
\and
\IEEEauthorblockN{Víctor Mayoral Vilches}
\IEEEauthorblockA{\textit{Alias Robotics} \\
Vitoria-Gasteiz, Spain\\
victor@aliasrobotics.com}
}

\maketitle

\begin{abstract}
Many recent approaches for vulnerability detection use large language models (LLMs) to classify functions as vulnerable
or benign, leaving developers to locate the exploitable statements. Furthermore, while these approaches have shown
performance improvements, they still fall short in distinguishing vulnerable from patched real-world functions.
This paper presents \apn{}, an agentic approach that detects exploitable statements in C/C++ functions.
In Phase~1, an LLM agent explores the function and its repository and identifies candidate statements.
In Phase~2, a second agent analyzes the control and data flow of each candidate and confirms it, rules it out,
or marks it uncertain.
We created a manually verified ground truth of exploitable statements for 200 pairs of the PrimeVul Paired test set
and ran \apn{} with the open-weight models Qwen 3.8 and DeepSeek V4.1 Flash and Anthropic's Sonnet 5.5 and Opus 5.5.
Phase~1 identifies ground-truth statements in 56--77\% of the vulnerable functions, and Phase~2 confirms
such a statement with an explanation accepted by an LLM judge in 26--73\%.
However, all models also confirm candidates in 23--61\% of the patched functions.
Across vulnerable and patched functions, the highest statement-level precision is 0.315 at a recall of 0.605
obtained with Opus 5.5 (xhigh).
At the function level, \apn{} with Opus 5.5 classifies 42.0\% of the pairs correctly, compared with 33.8\% reported for
the best previous approach.
\end{abstract}


\section{Introduction}
\label{sec:introduction}

To detect software vulnerabilities, static analysis tools encode vulnerability patterns as rules or as taint flows from sources
to sinks~\cite{livshits2005findingSecurityVulnerabilities,yamaguchi2015automaticInferenceSearch},
and code property graphs make such patterns queryable~\cite{yamaguchi2014modelingDiscoveringVulnerabilities}.
For C/C++, however, a recent comparison found that static analyzers produced tens of thousands of warnings
on programs containing about one hundred known bugs~\cite{hassler2025comparativeStudyFuzzers}.
Deep learning-based detectors instead learn vulnerability patterns from labeled functions or
statements~\cite{li2021vulnerabilityDetectionFinegrained,fu2022LineVulTransformerbasedLinelevel,steenhoek2024dataflowAnalysisInspiredDeep},
but their performance drops sharply on realistic, de-duplicated data, partly because they learn artifacts associated with noisy
labels rather than the vulnerabilities themselves~\cite{chakraborty2021deepLearningBased,croft2023dataQualitySoftware,chakraborty2024revisitingPerformanceDeep,risse2025topScoreWrong}.

Most recent approaches build on large language models (LLMs)~\cite{steenhoek2024errMachineVulnerability}. They
fine-tune LLMs to reason about vulnerabilities~\cite{wen2025boostingVulnerabilityDetection,weyssow2025R2VulLearningReason},
retrieve security knowledge from earlier vulnerabilities~\cite{du2026vulragEnhancingLLMbased,zhu2025VulInstructTeachingLLMs},
provide structural context from abstract syntax trees, control-flow and data-flow graphs, or code property
graphs~\cite{tang2026VulTriageTriplePathContext,lekssays2025LLMxCPG},
or let multiple agents debate or explore the code with tools~\cite{widyasari2025letTrialBegin,wang2026VulAgentHypothesisValidationDriven}.
On the PrimeVul Paired test set~\cite{ding2025vulnerabilityDetectionCode},
which pairs each vulnerable C/C++ function with its patched version, GPT-3.5 and GPT-4 perform close to random
guessing, and recent approaches still classify only a minority of the pairs correctly. For example,
VulTrial~\cite{widyasari2025letTrialBegin} classifies 81 of the 435 pairs (18.6\%) correctly, VulTriage~\cite{tang2026VulTriageTriplePathContext}
147 pairs (33.8\%), and VulInstruct~\cite{zhu2025VulInstructTeachingLLMs} 17.2\% of the pairs
when its explanation must also match the actual vulnerability.

Two limitations reduce the usefulness of these approaches for developers.
First, almost all of them classify a \emph{function} as vulnerable or benign. Even when they explain their decision,
as required by VulInstruct~\cite{zhu2025VulInstructTeachingLLMs} and CORRECT~\cite{li2025EverythingYouWantedKnow},
the developer still has to locate the statements that make the function exploitable.
Second, the few approaches that report vulnerable statements, such as LineVul~\cite{fu2022LineVulTransformerbasedLinelevel} and the
agentic pipeline of SecVulEval~\cite{ahmed2025SecVulEvalBenchmarkingLLMs}, take the lines changed by the fixing commit as
their ground truth. A fix, however, often adds a check before the exploitable statement and leaves the statement itself unchanged,
so this ground truth rewards identifying the changed lines rather than the exploitable statements.

%We therefore target detecting the statement(s) that an attacker can exploit, which we call \emph{exploitable sink}.
%The term sink stems from taint analysis, which describes a
%vulnerability as a flow of untrusted data from a source to a \emph{sink}, an unsafe use of the
%data~\cite{livshits2005findingSecurityVulnerabilities}. In C/C++ code, taint-style vulnerabilities are such flows that pass no
%sanitizer~\cite{yamaguchi2015automaticInferenceSearch}. We apply this term at statement level and define an
%\emph{exploitable sink} as a statement whose execution performs an operation that an attacker can exploit if the operation
%receives an unchecked value or runs in an unchecked state. For example, a \texttt{memcpy} without a size check or a
%statement that grants permissions to unauthorized users is such an exploitable sink. Further examples are an array write, a
%pointer dereference, a \texttt{free}, or an assertion, which covers the memory-safety and logic-level defects of C/C++ code.

We address these limitations with \apn{}, an agentic approach that detects \emph{exploitable statements} in C/C++ functions and explains
why they are exploitable. These statements perform the operations an attacker can exploit, such as an unchecked \texttt{memcpy}.
Following GPTLens~\cite{hu2023GPTLens}
and VulAgent~\cite{wang2026VulAgentHypothesisValidationDriven}, \apn{} splits the detection into two phases.
In Phase~1, an LLM agent explores the function and
its repository with file and text search tools and proposes candidates, each with its line number(s) and an explanation.
In Phase~2, a second agent investigates the control and data flow of every candidate and confirms it, rules it out, or
marks it uncertain. Based on these results, \apn{} classifies a function as vulnerable if Phase~2 confirms at least one candidate.
We implemented two versions of \apn{}, one with LangChain\footnote{\url{https://github.com/langchain-ai/langchain}} for open-weight models
and one with the Claude Code Agent SDK\footnote{\url{https://github.com/anthropics/claude-agent-sdk-python}} for Anthropic's frontier models.

For the evaluation of \apn{}, we created a ground truth for 200 randomly selected pairs from the PrimeVul Paired test set~\cite{ding2025vulnerabilityDetectionCode}.
We prompted an LLM to identify exploitable statements using the CVE description, the commit message, and the diff, and
manually verified the results for every vulnerable function.
A confirmed verdict counts as a true positive only if its line number matches a ground-truth statement and an LLM judge accepts its explanation.
Every confirmed verdict on a patched function counts as a false positive.
We ran \apn{} with Qwen 3.8, DeepSeek V4.1 Flash, Sonnet 5.5, and Opus 5.5 to answer three research questions:

\begin{description}
    \item[RQ1:] What is the performance of \apn{} in finding candidates for exploitable statements in C/C++ functions?
    \item[RQ2:] What is the performance of \apn{} in confirming the exploitable statements of C/C++ functions?
    \item[RQ3:] What is the performance of \apn{} in distinguishing vulnerable from patched C/C++ functions?
\end{description}

Phase~1 identifies candidates that match at least one ground-truth statement in 56--77\% of the vulnerable functions (RQ1).
Phase~2 confirms at least one ground-truth statement with an accepted explanation in 26--73\% of the vulnerable functions,
with Opus 5.5 performing best, followed by DeepSeek V4.1, Sonnet 5.5, and Qwen 3.8.
At the same time, all models also confirm candidates in 23--61\% of the patched functions, hence
the verdicts alone separate vulnerable from patched code only partly (RQ2).
With Opus 5.5, the classification based on confirmed Phase~2 verdicts exceeds the reported results of previous approaches:
\apn{} correctly classifies 42\% of the pairs, compared with 33.8\% for VulTriage and 18.6\% for VulTrial (RQ3).
Under the lenient CORRECT criterion, which also requires an accepted explanation, \apn{} with Opus 5.5 classifies 75.5\% of the pairs correctly,
compared with 17.2\% for VulInstruct (RQ3).
%Under the lenient CORRECT criterion, it reaches 75.5\%, compared with 17.2\% for VulInstruct (RQ3).

In summary, this paper makes the following contributions:
\begin{itemize}
    \item \apn{}, an agentic approach that detects exploitable statements in C/C++ functions in two phases and
    derives the function-level classification from the confirmed statements.
    \item Two implementations to run \apn{} with open-weight models and with Anthropic's frontier models.
    \item A ground truth of manually verified exploitable statements for 200 pairs of the PrimeVul Paired test set.
    \item An evaluation of \apn{} with four LLMs and different reasoning efforts at statement and function level.
    \item A replication package with all tools, the ground truth, the instructions, and all results.\footnote{\url{https://doi.org/10.5281/zenodo.23881951}}
\end{itemize}

%The remainder of the paper is
% Requires: \bibliography{relatedWork} (verified entries in relatedWork.bib)
% Preprint-only entries (arXiv) are marked note={preprint} in relatedWork.bib
\section{Related Work}
\label{sec:related}

We review related work on LLM-based vulnerability detectors, followed by approaches to detect vulnerabilities at statement or line level.

\paragraph{LLM-based Vulnerability Detection}
Ding \etal introduced PrimeVul~\cite{ding2025vulnerabilityDetectionCode}, a vulnerability dataset that
merges and relabels existing datasets such as Big-Vul~\cite{fan2020cCCode} and
DiverseVul~\cite{chen2023DiverseVulNewVulnerable}, with more accurate labels, de-duplication,
a chronological split, and a paired test set of vulnerable and patched functions. They evaluated code
language models and LLMs on it, among them GPT-3.5 and GPT-4, and showed that these models perform
close to random guessing on the paired test set. Other studies~\cite{steenhoek2024errMachineVulnerability,du2026vulragEnhancingLLMbased}
confirm that LLMs hardly distinguish vulnerable from patched code.

Most recent approaches therefore add knowledge or context about the function under analysis.
The evaluation framework CORRECT~\cite{li2025EverythingYouWantedKnow} supplies context on callees, types, and globals, enabling LLMs to reach an accuracy of up to 0.67.
An LLM judge checks whether their explanations match the ground truth.
VulInstruct~\cite{zhu2025VulInstructTeachingLLMs} retrieves security specifications extracted from earlier patches
and reaches an F1-score of 0.45 and 17.2\% correct pairs on the PrimeVul Paired test set under the CORRECT criterion.
VulTriage~\cite{tang2026VulTriageTriplePathContext} verbalizes AST, CFG, and DFG information and generic CWE patterns
and classifies 147 of 435 pairs (33.8\%) correctly with GPT-4o, without checking the explanation.
LLMxCPG~\cite{lekssays2025LLMxCPG} fine-tunes one model to generate Joern\footnote{\url{https://joern.io/}} queries
that slice the code property graph and another to classify the slices, reaching an F1-score of 0.62 for memory-safety CWEs on 40 PrimeVul pairs.
Other approaches fine-tune LLMs to reason about vulnerabilities.
ReVD~\cite{wen2025boostingVulnerabilityDetection} reaches a VP-Score (correct minus reversed pairs) of 16.1\% on PrimeVul,
and VULPO~\cite{li2025VULPOContextAwareVulnerability} correctly classifies 22.6\% of pairs from PrimeVul and SecVulEval.
%R2Vul~\cite{weyssow2025R2VulLearningReason} was evaluated without patched functions.

Agentic detectors divide the task among several agents. VulTrial~\cite{widyasari2025letTrialBegin} lets
four GPT-4o agents debate a code snippet and classifies 81 of 435 (18.6\%) pairs correctly.
Like \apn{}, GPTLens~\cite{hu2023GPTLens} separates candidate identification by an auditor from assessment by a critic.
The critic ranks candidates, and the approach correctly classifies 44 of 435 pairs (10.1\%) with GPT-4o~\cite{widyasari2025letTrialBegin}.
Phoenix~\cite{wang2026securityRelativeTrainingFree} reports 64.4\% correct pairs with 7--14B models, but
synthesizes its behavioral contract from both functions of the same test pair, their CVE description, and the commit message,
hence the ground truth enters the specification.
VulnAgent-R2~\cite{meng2026VulnAgentR2EvidenceCalibratedMultiAgent} reaches an F1-score of 0.39 on
the imbalanced PrimeVul test set (3\% vulnerable functions) against 0.70 on DiverseVul. Further agentic detectors
validate hypotheses against project context (VulAgent~\cite{wang2026VulAgentHypothesisValidationDriven}),
build causal knowledge graphs (CLEAR~\cite{yun2026CLEARCausalContextBased}),
or orchestrate static analysis (Argus~\cite{liang2026argusReorchestratingStatic}).
Unlike \apn{}, most of these approaches classify functions without locating the exploitable statements,
and they rely on fine-tuning, a retrieval corpus, or context extracted in advance.

\paragraph{Statement-Level Detection}
Static application security testing (SAST) tools, such as Flawfinder~\cite{wheeler2001Flawfinder}, Cppcheck~\cite{marjamaki2007Cppcheck}, Semgrep~\cite{semgrep2020Semgrep},
Infer~\cite{calcagno2015MovingFastSoftware}, the Clang Static Analyzer~\cite{llvm2020ClangStaticAnalyzer}, and CodeQL~\cite{avgustinov2016QLObjectoriented},
report warnings on statements based on rules or semantic analyses. On 112 memory-safety
bugs in C/C++, the best of five analyzers, CodeQL, detected 31 bugs, but only 0.24\% of its reports corresponded to a known bug,
and most analyzers had to be integrated into the build~\cite{hassler2025comparativeStudyFuzzers}.
Learning-based detectors use trained models to identify statements.
IVDetect~\cite{li2021vulnerabilityDetectionFinegrained} identifies statements that explain a function-level prediction,
while LineVul~\cite{fu2022LineVulTransformerbasedLinelevel} ranks lines within functions predicted as vulnerable.
LineVD~\cite{hin2022LineVDStatementlevelVulnerability} classifies statements with a graph neural network,
and SLVD-CC~\cite{chen2026SLVDCCEnhancingStatementlevel} classifies nodes in value-flow slices of compilable code.
Their ground truth consists of changed lines or synthetic test-case annotations.
IVDetect and LineVul evaluate localization only on vulnerable functions, while SLVD-CC was evaluated only on a synthetic dataset.
SecVulEval~\cite{ahmed2025SecVulEvalBenchmarkingLLMs} comes with a five-agent pipeline whose context agent extracts the definitions of the symbols a function uses.
A vulnerable function counts as detected if the pipeline names, with a correct explanation, at least one statement that the fixing commit deletes or adds or a sink statement.
The best configuration reaches an F1-score of 0.24 on 1,000 vulnerable and patched functions.
LLM-based localization has also been studied from crash evidence (T2L~\cite{xi2025traceLineLLM})
and for coding agents at repository level (VulnGym~\cite{ji2026VulnGymBenchmarkingCoding} and the
Vulnerability Localization Benchmark~\cite{priyanshu2026vulnerabilityLocalizationBenchmark}).
%The latter two give a single off-the-shelf agent the advisory or CWE description, so the agent localizes a known vulnerability,
%whereas \apn{} detects and verifies exploitable statements without such information.

\apn{} differs from these detectors in four respects.
First, it needs no training and no compilation, and its candidates
are its primary output rather than an explanation of a function-level decision.
Second, our ground truth consists of manually verified exploitable statements instead of the changed lines.
Third, the agents of \apn{} retrieve the context with tools, and Phase~2 assigns a verdict to every candidate.
Fourth, we count true and false positives per statement over the vulnerable and the patched
functions, and a true positive also needs an explanation accepted by an LLM judge.

\section{\apn{} Approach}
\label{sec:approach}

\begin{figure*}[t]
\centering
\resizebox{0.8\textwidth}{!}{%
% Overview of SBVulDetect (Figure~\ref{fig:overview}), included by approach.tex.
\begin{tikzpicture}[
  font=\footnotesize, >={Stealth[length=2mm]},
  phase/.style={draw, thick, rounded corners=3pt, fill=blue!8, minimum width=24mm, minimum height=12mm, align=center},
  optional/.style={phase, densely dashed},
  input/.style={draw, fill=gray!10, minimum width=20mm, minimum height=8mm, align=center, inner sep=2pt},
  doc/.style={draw, fill=white, minimum width=20mm, minimum height=6mm, align=center, inner sep=1pt},
  tool/.style={draw, rounded corners=2pt, fill=orange!12, minimum width=23mm, minimum height=8mm, align=center, inner sep=2pt},
  result/.style={align=center},
  flow/.style={->, thick},
  edge label/.style={font=\scriptsize\itshape, fill=white, inner sep=1pt},
]
% ---- inputs: the target function (taken from the code repository) and the instructions
\node[input] (fun) at (0,0) {target\\function};
\node[doc] (ins1) at (0,1.05) {instructions};
\node[draw, cylinder, shape border rotate=90, aspect=0.18, fill=gray!10, minimum width=22mm, minimum height=13mm,
      align=center, inner sep=1pt] (repo) at (0,-1.9) {\rule{0pt}{2.2mm}code\\repository};
\draw[->] (repo.north) -- (fun.south);

% ---- Phases 1 and 2
\node[phase] (p1) at (3.4,0) {\textbf{Phase 1}\\Candidate\\Identification};
\node[phase] (p2) at (7.6,0) {\textbf{Phase 2}\\Candidate\\Verification};
\draw[flow] (fun) -- (p1);
\draw[thick] (ins1.east) -- (p2.north |- ins1.east);
\draw[flow] (p1.north |- ins1.east) -- (p1.north);
\draw[flow] (p2.north |- ins1.east) -- (p2.north);
\draw[flow] (p1) -- node[edge label, below=1pt] {candidates} (p2);

% ---- output of Phase 2 and the rule-based classification below it (verdicts only)
\node[input] (ver) at (11.4,0) {verdicts\\{\scriptsize line, verdict, explanation}};
\node[optional] (p3) at (11.4,-1.85) {\textbf{Classification}\\{\scriptsize(rule-based)}};
\node[result, anchor=west] (res) at ([xshift=4mm]p3.east) {vulnerable /\\benign};
\draw[flow] (p2) -- (ver);
\draw[flow] (ver) -- (p3);
\draw[flow] (p3) -- (res);

% ---- tools on the code repository
\node[tool, minimum width=58mm] (tools) at (5.5,-1.85) {filesystem and text search tools};
\draw[->, thick] (p1.south) -- node[edge label, right] {tool calls} (p1.south |- tools.north);
\draw[->, thick] (p2.south) -- node[edge label, right] {tool calls} (p2.south |- tools.north);
\draw[->] (tools.west) -- (tools.west -| repo.east);
\end{tikzpicture}}
\caption{Overview of \apn{}. Phase~1 identifies exploitable statements. Phase~2 gathers evidence with tools to confirm them, rule them out, or mark them uncertain.
Based on the Phase~2 verdicts, Classification labels a function vulnerable or benign.}
\label{fig:overview}
\end{figure*}
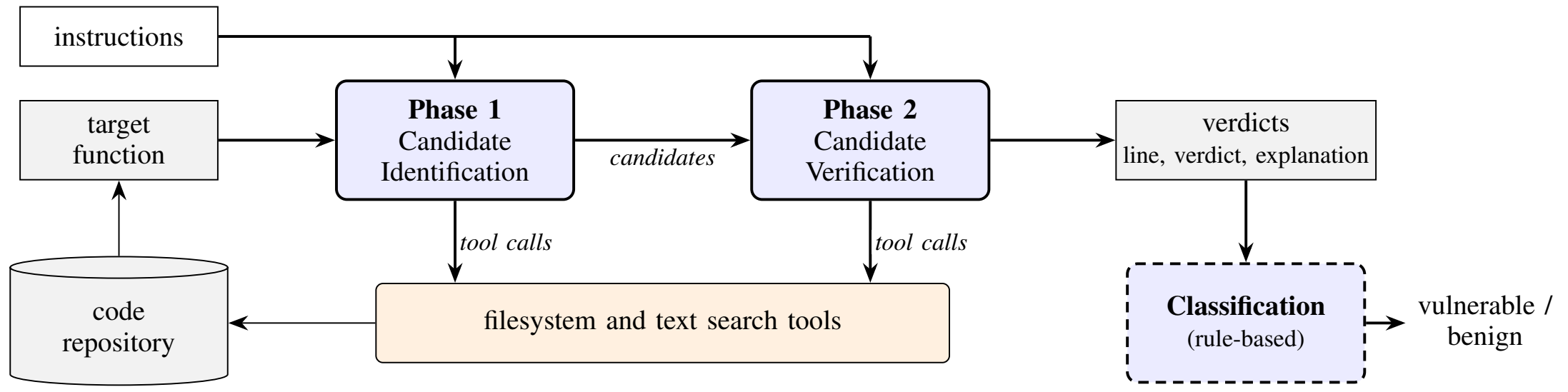

%In this section, we introduce \apn{}, our approach to detect exploitable statements in potentially vulnerable C/C++ functions.
\figref{fig:overview} shows an overview of \apn{}, which consists of two analysis phases followed by an optional
classification step. In Phase 1, \apn{} analyzes the source code of a given function, identifies potentially exploitable statements,
and outputs them each with the line number and a short explanation.
Phase 2 verifies each candidate from Phase 1 by a detailed analysis of its control and data flow and outputs a
verdict for it (confirmed, ruled out, or uncertain) together with a short explanation.
Based on the verdicts, \apn{} derives the classification of the function as vulnerable or benign.
The split into two phases
is motivated by previous work, such as GPTLens~\cite{hu2023GPTLens} and VulAgent~\cite{wang2026VulAgentHypothesisValidationDriven},
that found such a split to reduce the number of false positives while improving precision of the detection.
Phases 1 and 2 run as two separate agents that receive the same system prompt but share no conversation history.
Both agents investigate the code with filesystem and text search tools that list directories, find files by name,
search file contents, and read exact line ranges.
In the following, we describe each phase in more detail.

\subsection{Phase 1 - Candidate Identification}
\label{sec:approach:phase1}

\begin{lstlisting}[float=htb, caption={Excerpt of the system's prompt general and Phase 1 instructions for identifying exploitable statements.}, label=lst:instr-phase1]
You are a highly specialized C/C++ vulnerability detection expert [...] Your task is to determine whether a given C/C++ function contains an exploitable vulnerability.

## Phase 1 - Explore & Hunt for Candidate Statements
[...] Start by reading and reasoning about the target function's source above. Analyze:
- Structure & call graph: find its callers and see what it actually does with the arguments [...]
- Taint sources & sinks: search for how the target function's parameters/return value are produced upstream [...] and search for dangerous sink calls [...]

Combine tool findings with direct code reading [...]. In particular pay attention to:
- Memory operations: malloc, realloc, free, memcpy, [...]
- Integer arithmetic used as a size or index [...]
- Stale references across a side-effecting call: [...]

The line you report for each candidate must be the statement whose execution actually performs the unsafe operation [...] Output a JSON array - one object per candidate:
[{"line": <file line number as integer>, "code": <brief code snippet>, "reason": <...>}]
\end{lstlisting}

\lstref{lst:instr-phase1} shows an excerpt of the system prompt's general and Phase~1 instructions.
The agent receives the target function's name, source code, and file path. It starts with the function and extends its analysis to callers and callees.
For each caller, it analyzes whether the arguments can be derived from untrusted input,
such as network data or file contents, and are thus controllable by an attacker. For each callee, it traces where the
arguments come from and where the return value is used, searching for exploitable statements,
such as memory copies and command execution, that these values may reach without a check.
Besides the memory-safety patterns, such as unchecked copies,
integer arithmetic on sizes and indices, use-after-free, uninitialized reads, and format strings, the instructions list
logic-level patterns. These are missing validation, assertions that
attacker-controlled input can trigger, type confusion, missing authorization or completeness checks, and stale references.
The agent is also instructed to report potentially exploitable statements outside these patterns.
It stops when it finds no further candidates or reaches the tool-call limit.

% Idea: we might add a running example, but I am not sure whether this is really needed to make things clear

\subsection{Phase 2 - Candidate Verification}
\label{sec:approach:phase2}

\begin{lstlisting}[float=t, caption={Excerpt of the system's prompt Phase 2 instructions for verifying the candidates from Phase~1.}, label=lst:instr-phase2]
## Phase 2 - Confirm Each Candidate Statement
The orchestrator will provide the candidate list from Phase 1 and ask you to confirm or rule out each one using the available tools. For each candidate, investigate by:
- Reading the exact lines around the candidate again in context (`read_lines`) [...]
- Searching for other call sites of the function [...]
- [...]
A verdict must rest on evidence you actually gathered, not an assumption [...] If you haven't verified that assumption, the candidate is `uncertain` [...]
For each candidate record whether a vulnerability is confirmed, ruled out, or uncertain.
Stop investigating a candidate as soon as the evidence is conclusive. Output a JSON array [...]
\end{lstlisting}

\lstref{lst:instr-phase2} shows an excerpt of the Phase~2 instructions.
The agent receives the candidate list from Phase~1 and assigns each candidate the verdict
\emph{confirmed}, \emph{ruled out}, or \emph{uncertain}.
While Phase~1 explores the code broadly, Phase~2 seeks evidence to determine whether a candidate is exploitable.
The agent examines the surrounding control and data flow, guards, and argument validation in callers.
It reads type and macro definitions to establish relevant buffer sizes and layouts, and traces allocations and bounds checks across adjacent code.

Every verdict must rest on evidence gathered with tools. If that evidence neither confirms nor rules out a candidate,
the agent marks it \emph{uncertain}. For a \texttt{memcpy}, for example, the agent must trace the origin of the copy size
and explain how an attacker could influence it. For a logic-level candidate, it must cite a specific control or data flow path.
An empty tool result alone provides no evidence in either direction.

The optional classification step labels a function as vulnerable if Phase~2 confirms at least one candidate, and as benign otherwise.
Uncertain verdicts do not count as positive predictions because Phase~2 could not establish exploitability.
%Counting them as vulnerable, would favor recall at the cost of precision.
This step adds no analysis of its own, in contrast to detectors that predict the class directly,
such as VulTrial~\cite{widyasari2025letTrialBegin} and VulTriage \cite{tang2026VulTriageTriplePathContext},
or that let an LLM judge the code against a specification, such as Phoenix~\cite{wang2026securityRelativeTrainingFree}.

%Prior agentic detectors gather context through retrieval, project graphs, or specialized
%agents~\cite{wang2026VulAgentHypothesisValidationDriven,meng2026VulnAgentR2EvidenceCalibratedMultiAgent,liang2026argusReorchestratingStatic}.
%To the best of our knowledge, the combination of plain filesystem and text search tools, an explicit analysis scope, and a per-phase
%reasoning effort in one loop is not described in these works. It allows us to measure, on the same pairs and with the same
%instructions and tools, how much the reasoning effort contributes to detection quality.

\section{Data Preparation}
\label{sec:dataset}

In this section, we first describe the selection of the test dataset and then the curation of the ground truth on statement level.

\subsection{Test Dataset}
We used the PrimeVul Paired test set~\cite{ding2025vulnerabilityDetectionCode}, which
contains 435 pairs of a vulnerable function and its patched version.
We selected this dataset because it is de-duplicated and
split chronologically, and its labels reach an accuracy comparable to human-verified benchmarks.
Furthermore, previous studies, such as VulTrial~\cite{widyasari2025letTrialBegin} and VulTriage~\cite{tang2026VulTriageTriplePathContext},
evaluated their detectors on it, allowing comparison with published results.

For each pair, our checkout tool fetched a shallow copy of the repository containing only the parent and
fixing commits. The checkout succeeded for 386 pairs and failed for 49 pairs, mainly because the
repository had moved from GitHub to another git server. We manually fixed the URLs of 28 pairs and skipped the
remaining 21 pairs. From the resulting 414 pairs, we randomly selected 200 to keep the runtime and cost of evaluating several configurations and LLMs manageable. To prevent a few large projects from dominating the sample, we selected at most 10 pairs
per project. Without this limit, TensorFlow (72 pairs), the Linux kernel (68 pairs), and Vim (38 pairs) would make up
43\% of the 414 pairs. In our sample, the largest project contributes 9 pairs.

\begin{table}[t]
\centering
\caption{Descriptive statistics of the 200 sampled pairs from 104 projects.}
\label{tab:dataset}
\begin{tabular}{p{4.2cm}r}
\hline
Most frequent CWEs & CWE-787 (38), CWE-125 (27), \\
 & CWE-703 (16), CWE-476 (14) \\
\#CVEs from 2021 or 2022 & 134 \\
\#.c/.cpp/.cc/.h files & 162 / 19 / 18 / 1 \\
LOC per vulnerable function (median/mean/max) & 96 / 211 / 2,411 \\
\hline
\end{tabular}
\end{table}

\tabref{tab:dataset} shows descriptive statistics of the sampled 200 pairs.
Our dataset covers 104 C/C++ projects, 45 types of common weaknesses (CWEs), most of
them contained in .c files, with a median number of lines of code (LOC) of 96 and maximum 2,411.
About two thirds of the vulnerabilities (CVEs) were reported in 2021 and 2022 and the oldest date back to 2008.

\subsection{Ground Truth}
For statement-level evaluation, we annotated the vulnerable functions in the 200 pairs with their exploitable statements.
Our helper tool prompts an LLM to locate statements that perform the exploited operation, such as an out-of-bounds \texttt{memcpy} or array access,
an unsafe \texttt{free}, an unchecked size computation, or a branch taken without authorization. Unlike \apn{}, the tool receives the CWE type, the CVE number and description, the commit message,
the names of the vulnerable function and its source file, the diff, and the source code
annotated with line numbers. For each function, the tool outputs the line numbers of the exploitable statements
and a rationale explaining why they are exploitable.
A statement spanning several lines is listed with all of its lines.
We ran this tool with Opus 5.5 and high reasoning effort.

Two authors then individually verified the output for the first 20 functions against the source code, the diff, and the commit message.
In a meeting, they discussed every function for which one of them questioned the output and decided on each statement.
Next, each author individually verified the remaining 180 functions, and in a second meeting
they discussed all differences until they reached consensus. In total, the authors revised the
ground truth of six functions, mainly by replacing a guard with the statement that performs the exploited operation.
They removed four pairs whose vulnerable function contains no exploitable statement, because the vulnerability is a missing statement
whose effect becomes visible only in other functions, and replaced them with randomly drawn pairs for which the
previously described process was repeated.
%For example, the
%vulnerable function of pair 195308 misses an entry in the initialization of a \texttt{syscall\_blocklist}.

\begin{table}[t]
\centering
\caption{Number of exploitable statements per vulnerable function.}
\label{tab:gt}
\begin{tabular}{lcccccccc}
\hline
Statements & 1 & 2 & 3 & 4 & 5 & 6 & 8--9 & 10--14 \\
Functions & 118 & 46 & 14 & 11 & 4 & 2 & 2 & 3 \\
\hline
\end{tabular}
\end{table}

\tabref{tab:gt} shows that the ground truth comprises 380 statements on 460 lines.
Only 37\% of the 380 ground-truth statements are modified by the patch.
In 113 of the 200 vulnerable functions (56\%), the patch leaves all ground-truth statements unchanged,
and 58 patches only insert code, typically a check before an unchanged exploitable statement.
Using lines added or deleted by the patch as ground truth would therefore reward identifying the fix rather than the exploitable statement.

\section{Experimental Setup}

This section presents the experimental setup used to answer the research questions. First, we describe
the two implementations of \apn{} and their configurations used in our runs followed by the metrics used to
evaluate it.

\subsection{Agents Implementation}
\label{sec:setup:harness}
We implemented two versions of \apn{}, one with LangChain (LC) and one with the Claude Code Agent SDK (CC),
which use the same instructions and equivalent read-only tools on the same export of the source code.

\subsubsection{LangChain}
The implementation with LangChain uses a custom LLM-tool loop to handle timeouts, context length exceed errors, and malformed tool calls,
and to enforce the tool-call limit. These errors particularly affected open-weight models in our test runs.
When an error is encountered, the agent returns the results gathered so far.
For instance, if an LLM call with reasoning effort medium times out, it retries the call with reasoning effort \emph{none}.
The filesystem\footnote{\url{https://github.com/modelcontextprotocol/servers/tree/main/src/filesystem}}
and ripgrep\footnote{\url{https://github.com/mcollina/mcp-ripgrep}} MCP servers provide the tools for directory listing, file-name search, and
content search. In addition, \apn{} provides a tool to read exact line ranges from source files.

With this setup, we used the open-weight models Qwen3.8-27B-FP8 (Qwen 3.8), served with vLLM 0.27.1,
and DeepSeek V4.1 Flash 1017 (DeepSeek V4.1), accessed via DeepSeek's OpenAI-compatible API.
%In addition, we the commercial Sonnet 5.5 model by Anthropic accessed via an API key.
%We also performed test runs with Qwen3.6-27B and other open weight models, such as ..., of which Qwen3.8 showed
%the best performance.
At the time of our experiments (August to October 2026), Qwen 3.8 (27B parameters, 256k-token context) was among the strongest open-weight models of its size
on the Artificial Analysis Intelligence Index (score 34)\footnote{\url{https://artificialanalysis.ai/models/qwen3-8-27b}},
and DeepSeek V4.1 (552B parameters, 16B active, 1M-token context) performed comparably to large commercial models (score 45)
at a lower cost per task\footnote{\url{https://artificialanalysis.ai/models/deepseek-v4-1-flash-1017}}.
We performed all experiments in a virtual machine with an 8-core AMD CPU, 62\,GB of main memory, an
NVIDIA H100 NVL with 94\,GB, and Ubuntu 22.04.5 LTS. vLLM served Qwen 3.8 with the maximum context length,
prefix caching, and the recommended settings for the model\footnote{\url{https://recipes.vllm.ai/Qwen/Qwen3.8-27B?variant=fp8}}.
We used greedy decoding (temperature 0, seed 42) with top-k 15, min-p 0.02, and a repetition penalty of 1.02.
Qwen 3.8 does not bound its thinking, therefore we capped \emph{medium} reasoning effort at 4,096 thinking tokens.
Based on several test runs on our hardware, we set the timeouts of an LLM call to 90\,s and wrap-up call to 180\,s,
the timeout of a tool call to 30\,s, and the maximum number of tool calls to 80 per phase.

DeepSeek V4.1 offers the reasoning efforts low, high, and max, and \emph{none} disables thinking.
We limited the output of a DeepSeek call, including thinking, to 16,000 tokens,
because very long thinking exceeds the call timeout and increases the cost.
We kept the same timeouts as for Qwen 3.8, except for the runs with max effort for which we increased the
timeouts to 300\,s and 600\,s respectively.
DeepSeek's API ignores the seed and, in thinking mode, the temperature, so runs with DeepSeek V4.1 are not
exactly reproducible.

\subsubsection{Claude Code Agent SDK}
The agents implemented with the Claude Code Agent SDK run in a Docker container to better restrict internet and filesystem access.
The agents see only the vulnerable or patched version of the source code exported from the code repository.
Each agent runs as a separate Claude Code session, and Claude Code's own loop executes the tool calls.
Out instructions replace Claude Code's default system prompt and differ from those of the LangChain implementation only in two respects.
First, they name Claude Code's tools \texttt{Read}, \texttt{Grep}, and \texttt{Glob}.
Second, the agent passes the output format as a JSON schema instead of in a wrap-up prompt.
%In contrast to the LangChain implementation, we did not implement our own LLM-tool loop
We limited a session to 80 turns, each of which may contain several tool calls,
and to 1,800\,s per function and phase.
%No session reached these limits.
%Claude Code offers the reasoning efforts low, medium, high, xhigh, and max, but no option without thinking.

\subsection{Metrics}
\label{sec:setup:metrics}
We evaluated the output of \apn{} in the two phases with statement-level and function-level metrics against
our ground truth. For Phase~2, we additionally evaluated the explanation of each verdict against the ground
truth with an LLM judge.

\subsubsection{LLM Judge}
\label{sec:setup:judge}
As found in previous work, such as CORRECT~\cite{li2025EverythingYouWantedKnow} and VulInstruct~\cite{zhu2025VulInstructTeachingLLMs},
matching a ground-truth statement does not establish that a detector identified the vulnerability for the right reason.
We therefore use an LLM judge to verify whether the explanation of a confirmed verdict in Phase~2 matches the reasoning in the ground truth.
We use Sonnet 5.5 with high effort as judge for all models.
To check for self-preference, we also judged the verdicts of Sonnet 5.5 and Opus 5.5 with DeepSeek V4.1, which changes their true positives (TPs) by at most four
and their hit rates by at most one percentage point.
The LLM judge runs once per function with at least one confirmed or uncertain verdict, without tools and without the source code.
It receives the ground-truth line numbers and rationale, together with the line number and explanation of each confirmed or uncertain verdict.
It then checks whether each explanation matches the root cause, independently of the line number.
For a patched function, the judge instead decides whether the explanation points at the vulnerability that the patch fixed.
Ruled-out verdicts claim no vulnerability and are not judged.

\subsubsection{Statement-Level Metrics}
As in previous work by Hin \etal~\cite{hin2022LineVDStatementlevelVulnerability} and Chen \etal~\cite{chen2026SLVDCCEnhancingStatementlevel},
each ground-truth statement is one instance. Note, a statement can span multiple lines of code and every remaining line in the function is a separate instance.
In Phase~1, a candidate statement is a true positive (TP) if it is contained in the ground truth, and a false negative (FN) otherwise.
A candidate statement not contained in the ground-truth is a false positive (FP), and every remaining line in the function is a true negative (TN).
In a patched function, every candidate is an FP and every other line in the function is a TN.

In Phase~2, a statement in a vulnerable function is a TP if it has a confirmed verdict contained in the ground truth and the LLM judge
accepts the verdict's explanation, and an FN otherwise. Every other confirmed verdict is an FP, including one with a rejected explanation.
In the patched function, each confirmed verdict is an FP.
From these counts, we compute precision (P), recall (R), F1-score, and the Matthews correlation coefficient (MCC).
%With an average of 211 lines but only 1.9 ground-truth statements per vulnerable function, TNs dominate,
%and MCC is close to the geometric mean of precision and recall.
%The rows \emph{all} of the tables cover the vulnerable and patched functions together.

\subsubsection{Function-Level Metrics}
The \emph{hit rate} (Hit) is the share of vulnerable functions with at least one TP.
Cand is the average number of Phase~1 candidates per function. In Phase~2, $\geq 1$ Conf is the share of functions with at least
one confirmed verdict, whether or not it matches the ground truth.
For the classification, we use the pair-wise metrics of PrimeVul~\cite{ding2025vulnerabilityDetectionCode}.
A pair is \emph{pair-correct} (P-C) if the vulnerable function is classified as vulnerable and the patched function as benign,
\emph{pair-vulnerable} (P-V) if both are classified as vulnerable, \emph{pair-benign} (P-B) if both are classified as benign, and
\emph{pair-reversed} (P-R) otherwise.

Following CORRECT~\cite{li2025EverythingYouWantedKnow} and SecVulEval~\cite{ahmed2025SecVulEvalBenchmarkingLLMs},
we also check whether a classification is correct for the \emph{right reason} (RR).
A vulnerable function classified as vulnerable counts as a TP only if it has a Phase~2 TP. Otherwise, it counts as an FN.
Every patched function classified as vulnerable is an FP.
A pair is P-C$_{RR}$ if it is P-C and its vulnerable function has a Phase~2 TP.
For a comparison with CORRECT and VulInstruct~\cite{zhu2025VulInstructTeachingLLMs}, we also report the metrics under the lenient \emph{CORRECT criterion} (C).
A vulnerable function classified as vulnerable is a TP if the judge accepts at least one of its verdicts regardless of its line number,
and a patched function classified as vulnerable is an FP only if at least one of its verdicts points at the fixed vulnerability.
Our right-reason criterion (RR) differs from CORRECT, which checks explanations against the CVE description and diff rather than a verified ground truth.
It differs from SecVulEval, which accepts at most two unrelated statements per function.
One TP suffices for a function to count as correct for the right reason, while the statement-level metrics also penalize false confirmed verdicts.

We count the functions with a hard timeout (HT), \ie an LLM call that failed even after the retry, and with a soft timeout (ST), \ie a call that succeeded when retried with reasoning effort \emph{none}.
Finally, we measured the runtime, tool calls, tokens, and estimated cost per function. We provide the latter two measurements in the replication package.

\section{Results}
\label{sec:results}

In this section, we present the results to answer the research questions. All results are provided in the replication package.

\subsection{Results and Answer to RQ1}
\label{sec:rq1}

\tabref{tab:rq1-phase1} shows the Phase~1 results for the best configurations obtained with
the LangChain (LC) and Claude Code (CC) agents. For Phase~1, we aim at a high hit rate and at a high recall
for vulnerable functions.

\begin{table*}[t]
\centering
\caption{Best Phase~1 results on the 200 pairs per model and reasoning effort for the vulnerable (vul) and patched (pat) functions.}
\label{tab:rq1-phase1}
\footnotesize
\setlength{\tabcolsep}{5pt}
\begin{tabular}{l | l | l | r | rrr | rrrr | r | rrr}
\hline
Model & Effort & Label & Hit & TP & FP & FN & P & R & F1 & MCC & Cand & HT/ST & TC & Time \\
\hline
Qwen $3.8_{LC}$ & none & vul & 56\% & 173 & 643 & 207 & 0.212 & 0.455 & 0.289 & 0.302 & 4.12 & 2 / 0 & 32.6 & 103\,s \\
 & & pat & - & - & \textbf{772} & - & - & - & - & - & 3.86 & 1 / 0 & 36.4 & 109\,s \\
\cline{3-15}
 & & all & - & 173 & 1415 & 207 & 0.109 & 0.455 & 0.176 & 0.216 & 3.99 & 3 / 0 & 34.5 & 106\,s \\
%\cline{2-15}
% & medium & vul & 55\% & 171 & 484 & 209 & 0.261 & 0.450 & 0.330 & 0.335 & 3.29 & 0 / 10 & 20.3 & 399\,s \\
% & & pat & - & - & 722 & - & - & - & - & - & 3.61 & 0 / 16 & 20.9 & 422\,s \\
%\cline{3-15}
% & & all & - & 171 & 1206 & 209 & 0.124 & 0.450 & 0.195 & 0.230 & - & - & - & - \\
\hline
DeepSeek $V4.1_{LC}$ & high & vul & 73\% & 233 & 646 & 147 & 0.265 & 0.613 & 0.370 & 0.395 & 4.45 & 0 / 7 & 31.7 & 168\,s \\
& & pat & - & - & 1020 & - & - & - & - & - & 5.10 & 0 / 9 & 35.0 & 202\,s \\
\cline{3-15}
 & & all & - & 233 & 1666 & 147 & 0.123 & 0.613 & 0.204 & 0.268 & 4.77 & 0 / 16 & 33.4 & 185\,s \\
\hline
Sonnet 5.5$_{CC}$ & medium & vul & 69\% & 216 & 1050 & 164 & 0.171 & 0.568 & 0.262 & 0.301 & 6.37 & - & 0.3 & 9\,s \\
& & pat & - & - & 1220 & - & - & - & - & - & 6.10 & - & 0.3 & 10\,s \\
\cline{3-15}
 & & all & - & 216 & 2270 & 164 & 0.087 & 0.568 & 0.151 & 0.214 & 6.24 & - & 0.3 & 9\,s \\
\hline
Opus 5.5$_{CC}$ & medium & vul & \textbf{77\%} & \textbf{250} & \textbf{610} & \textbf{130} & \textbf{0.291} & \textbf{0.658} & \textbf{0.403} & \textbf{0.430} & 4.33 & - & 1.8 & 18\,s \\
 & & pat & - & - & 957 & - & - & - & - & - & 4.79 & - & 2.7 & 23\,s \\
\cline{3-15}
 & & all & - & 250 & 1567 & 130 & \textbf{0.138} & \textbf{0.658} & \textbf{0.228} & \textbf{0.295} & 4.56 & - & 2.3 & 20\,s \\
\hline
\end{tabular}
\end{table*}

Qwen 3.8 (none) reaches a hit rate of 56\% (111 functions) and a recall of 0.455 with 173 TP statements. Although it is
the best Phase~1 configuration for Qwen 3.8, all other models outperform it in hit rate and recall.
DeepSeek V4.1 (high) reaches a hit rate of 73\% (145 functions) and a recall of 0.613.
Note that DeepSeek V4.1 offers no medium effort, and its high effort corresponds to the medium effort of Sonnet and Opus 5.5.
Sonnet 5.5 reaches a hit rate of 69\% (138 functions) and a recall of 0.568.
Its hit rate exceeds that of Qwen 3.8 but is slightly lower than that of DeepSeek V4.1.
Opus 5.5 (medium) performs best. It reaches the highest hit rate of 77\% (153 functions),
matches the most ground-truth statements (250, recall 0.658) and has the best precision (0.291),
F1-score (0.403), and MCC (0.430) with on average 4.3 candidates per vulnerable function.

Every model also proposes candidates in nearly all patched functions (3.9--6.1 per function), which are all FPs.
Considering both labels, precision therefore drops to 0.087--0.138. Qwen 3.8 has the fewest FPs on patched functions (772),
followed by Opus 5.5 (957).
Sonnet 5.5 and Opus 5.5 hardly use the tools, with on average 0.3 and 2.3 tool calls per function,
compared to 34.5 by Qwen 3.8 and 33.4 by DeepSeek V4.1.
With Claude Code, Sonnet 5.5 and Opus 5.5 process a function in 9--20\,s on average, compared to
106\,s and 185\,s for Qwen 3.8 and DeepSeek V4.1.

%Comparing the results of the four models and the different configurations,
%Opus 5.5 shows the best performance for Phase~1 followed by DeepSeek and then Sonnet 5.5.
%Its lead over DeepSeek is not significant (vulnerable functions with an exact ground-truth line among the
%candidates of only Opus 5.5 vs.\ only DeepSeek: 18 vs.\ 10, McNemar two-sided sign test $p = 0.18$). Both models
%outperform Sonnet 5.5 and Qwen3.8 in precision, recall, and MCC, and Qwen3.8 also in coverage
%(MCC over the vulnerable and benign functions together: Opus 5.5 0.271, DeepSeek 0.251, Sonnet 5.5 0.199, Qwen3.8 0.205--0.212).

\paragraph{Impact of the reasoning effort}
We ran the two versions of \apn{} with various reasoning efforts.
For Qwen 3.8, increasing the reasoning effort to medium does not improve the hit rate (55\%) and recall (0.450)
and improves precision only moderately (0.212 to 0.261 on the vulnerable functions), but
raises the runtime from 103\,s to 399\,s per function and causes 26 soft timeouts.

We tested DeepSeek V4.1 with none and max reasoning effort on a random sample of 49 pairs and compared it to the results
with high effort. Max effort raises the hit rate of DeepSeek V4.1 from 73\% to 78\% and the recall from 0.645 to 0.699, but
lowers the overall precision from 0.124 to 0.103 and uses 2.6 times as many tokens.
Increasing the reasoning effort therefore does not pay off for DeepSeek V4.1.
Disabling thinking lowers both the hit rate to 59\% and the precision to 0.049.

For Sonnet 5.5, on the same 49 pairs, xhigh effort raises the hit rate from 67\% to 78\%,
the recall from 0.602 to 0.763, and the overall precision from 0.071 to 0.110 compared to medium effort,
but at 6.7 times the runtime and 4.7 times the cost. Opus 5.5 with medium effort reaches
a similar hit rate (76\%) and recall (0.720) on these pairs with a higher overall precision (0.147),
at 41\% of the cost and a quarter of the runtime.
On the first 30 of these pairs, xhigh effort further raises the hit rate of Opus 5.5 from 80\% to 97\%
and the recall from 0.797 to 0.922, but lowers the overall precision at 4.7 times the runtime.
%Therefore, we defer a full Phase~1 run with xhigh effort to future work.

%Higher effort therefore mainly makes the candidate lists shorter and more precise; it clearly
%improves the coverage only for Sonnet 5.5, on a small sample.

\paragraph{Differences between CWE types}
We compare the hit rates per CWE type of the four configurations in \tabref{tab:rq1-phase1}.
%The results are provided in the replication package.
DeepSeek V4.1, Sonnet 5.5, and Opus 5.5 achieve their highest hit rates on out-of-bounds accesses (75 functions with CWE-787, -125, -119, or -120),
with 84\%, 76\%, and 87\%; Qwen 3.8 reaches 65--68\%.
NULL pointer dereferences (CWE-476, 14 functions) reach similar hit rates (71--86\% for DeepSeek V4.1, Sonnet 5.5, and Opus 5.5, 50--79\% for Qwen 3.8).
The hit rates for CWE-703, improper check or handling of exceptional
conditions (16 functions), are 31--38\% for Qwen 3.8, 69\% for DeepSeek V4.1, 50\% for Sonnet 5.5,
and 56\% for Opus 5.5. These defects center on missing or incorrect checks and involve more
ground-truth statements (2.3 per function vs.\ 1.9 for out-of-bounds accesses).
%The gap between out-of-bounds and CWE-703 is significant for Qwen3.8 with medium effort and Opus 5.5
%(Fisher's exact test, $p = 0.01$), but not for Qwen3.8 without reasoning ($p = 0.05$), DeepSeek ($p = 0.17$), and Sonnet 5.5 ($p = 0.06$).
%
Of the five resource management weaknesses
(CWE-400, -401, and -772), every configuration finds the same single function, and of the reachable assertions (CWE-617, 6 functions),
each configuration finds at most three, while all miss three.
The seven access-control and authentication weaknesses (CWE-269, -276, -284, -287, -288, and -522) are missed by
Qwen 3.8 and, except for one, by Sonnet 5.5, whereas Opus 5.5 finds five and DeepSeek V4.1 two.
Missing releases, checks, or permissions can make it difficult to identify a single exploitation point.
Overall, 28 of the 200 vulnerable functions are missed by every configuration.

\paragraph{Answer to RQ1}
In Phase~1, \apn{} identifies ground-truth statements in 56--77\% of the vulnerable
functions, with a recall of 0.45--0.66.
Opus 5.5 performs best with a hit rate of 77\%, followed by DeepSeek V4.1, Sonnet 5.5, and Qwen 3.8.
Higher reasoning effort raises hit rate and recall for DeepSeek V4.1, Sonnet 5.5, and Opus 5.5, with mixed effects on precision,
but at 2.6--6.7 times the tokens, cost, or runtime.
Patched functions receive about as many candidates as vulnerable ones, so the candidates require further verification.

\subsection{Results and Answer to RQ2}
\label{sec:rq2}

\begin{table*}[t]
\centering
\caption{Phase~2 results on the 200 pairs per model and reasoning effort for the vulnerable (vul) and patched (pat) functions.}
\label{tab:rq2-phase2}
\footnotesize
\setlength{\tabcolsep}{4pt}
\begin{tabular}{l | l | l| rr | rrr | rrrr | rrr}
\hline
Model & Effort & Label & Hit & $\geq 1$ Conf & TP & FP & FN & P & R & F1 & MCC & HT/ST & TC & Time \\
\hline
%Qwen$3.8_{LC}$ & none & vul & 26\% & 49\% & 89 & 186 & 291 & 0.324 & 0.234 & 0.272 & 0.270 & 0 / 0 & 31.4 & 100\,s \\
% & & pat & - & 31\% & - & 157 & - & - & - & - & - & - & 0 / 0 & 33.7 & 101\,s \\
%\cline{3-15}
% & & all & - & - & 89 & 343 & 291 & 0.206 & 0.234 & 0.219 & 0.216 & - & - & - & - \\
%\cline{2-15}
Qwen $3.8_{LC}$ & medium & vul & 26\% & 42\% & 92 & \textbf{120} & 288 & 0.434 & 0.242 & 0.311 & 0.320 & 0 / 10 & 14.7 & 197\,s \\
 & & pat & - & \textbf{23\%} & - & \textbf{104} & - & - & - & - & - & 0 / 9 & 14.1 & 178\,s \\
\cline{3-15}
 & & all & - & - & 92 & 224 & 288 & 0.291 & 0.242 & 0.264 & 0.263 & 0 / 19 & 14.4 & 188\,s \\
\hline
DeepSeek $V4.1_{LC}$ & high & vul & 57\% & 84\% & 177 & 297 & 203 & 0.373 & 0.466 & 0.415 & 0.411 & 0 / 2 & 25.0 & 117\,s \\
 & & pat & - & 61\% & - & 360 & - & - & - & - & - & 0 / 3 & 25.8 & 136\,s \\
\cline{3-15}
 & & all & - & - & 177 & 657 & 203 & 0.212 & 0.466 & 0.292 & 0.310 & 0 / 5 & 25.4 & 126\,s \\
\hline
Sonnet $5.5_{CC}$ & medium & vul & 43\% & 59\% & 139 & 197 & 241 & 0.414 & 0.366 & 0.388 & 0.384 & - & 5.2 & 27\,s \\
 & & pat & - & 37\% & - & 211 & - & - & - & - & - & - & 5.1 & 26\,s \\
\cline{3-15}
 & & all & - & - & 139 & 408 & 241 & 0.254 & 0.366 & 0.300 & 0.301 & - & 5.2 & 27\,s \\
\hline
Opus $5.5_{CC}$ & medium & vul & 66\% & 89\% & 215 & 213 & 165 & 0.502 & 0.566 & 0.532 & 0.529 & - & 4.6 & 31\,s \\
 & & pat & - & 52\% & - & 266 & - & - & - & - & - & - & 4.9 & 34\,s \\
\cline{3-15}
 & & all & - & - & 215 & 479 & 165 & 0.310 & 0.566 & 0.400 & 0.415 & - & 4.7 & 33\,s \\
\cline{2-15}
 & xhigh & vul & \textbf{73\%} & \textbf{94\%} & \textbf{230} & 220 & \textbf{150} & \textbf{0.511} & \textbf{0.605} & \textbf{0.554} & \textbf{0.552} & - & 13.2 & 107\,s \\
 & & pat & - & 54\% & - & 281 & - & - & - & - & - & - & 14.4 & 134\,s \\
\cline{3-15}
 & & all & - & - & 230 & 501 & 150 & \textbf{0.315} & \textbf{0.605} & \textbf{0.414} & \textbf{0.433} & - & 13.8 & 121\,s \\
\hline
\end{tabular}
\end{table*}

\tabref{tab:rq2-phase2} shows the best results of Phase~2 with the Phase~1 candidates from \tabref{tab:rq1-phase1}.
%Note, a statement in a vulnerable function is a TP if it has a confirmed verdict contained in the ground truth and the LLM judge
%accepts the verdict's explanation, and an FN otherwise.
We aim for high hit rate, and in addition to high recall also high precision.

Qwen 3.8 (medium) reaches a hit rate of only 26\% (52 functions),
although its Phase~1 candidates hit 56\% of the vulnerable functions.
It has the fewest FPs on patched functions of all configurations (104).
DeepSeek V4.1 reaches a hit rate of 57\% (113 functions).
However, it confirms candidates in 61\% of patched functions, the most of any model, compared with 84\% of vulnerable functions.
Sonnet 5.5 lies between Qwen 3.8 and DeepSeek V4.1 with a hit rate of 43\%.
Its hit rate of 66\% (131 functions) clearly exceeds DeepSeek V4.1's 57\%,
although its Phase~1 hit rate is only 4 percentage points higher (77\% vs.\ 73\%).
It outperforms DeepSeek V4.1 and Sonnet 5.5 in hit rate, precision, recall, F1-score, and MCC.
With xhigh effort, Opus 5.5 achieves the best results with a hit rate of 73\% (146 functions), a recall of 0.605,
a precision of 0.511, and an MCC of 0.552. However, Opus 5.5 confirms candidates in 52\% (medium)
and 54\% (xhigh) of the patched functions, with more FPs on patched functions (266 and 281) than Qwen 3.8 (104).

Considering both labels, overall precision drops by 0.12--0.20, relatively most for DeepSeek V4.1 (from 0.373 to 0.212),
leaving DeepSeek V4.1 and Sonnet 5.5 with similar MCC values (0.310 and 0.301), while Opus 5.5 stays ahead (0.415 and 0.433).
Compared to the rows \emph{all} of Phase~1, Phase~2 raises precision by a factor of 1.7 to 2.9 for all models while
recall drops by 0.05--0.21, least for Opus 5.5 (from 0.658 to 0.566 and 0.605) and most for Qwen 3.8 (from 0.455 to 0.242).

%Our evaluation is stricter than SecVulEval's, which counts per function and accepts one of several fix locations and up to two unrelated statements.
%The balanced sample also yields higher precision than expected on real code~\cite{arp2022dosDontsMachine,chakraborty2024revisitingPerformanceDeep},
%although distinguishing a vulnerable function from its patched version is harder than distinguishing it from unrelated code.

%Note, we also ran Phase~2 in a per-candidate mode, in which each candidate is analyzed in a separate tool loop, with Qwen3.8 and all six configurations on the first 97 pairs.
%It took 1.3 to 2.9 times as long and improved $\pm 0$ significantly only for CB with medium effort (28 vs.\ 18 functions, $p = 0.013$), so we used the default mode,
%which analyzes all candidates in one tool loop.

\paragraph{Impact of the reasoning effort}
For Qwen 3.8, using no reasoning results in almost the same hit rate (51 vs.\ 52 functions) and recall (0.234 vs.\ 0.242)
as medium effort, but in lower precision (0.324 vs.\ 0.434) and MCC (0.270 vs.\ 0.320).
For DeepSeek V4.1, max effort does not improve over high effort on the first 30 of the 49 sampled pairs,
but needs three times the runtime and 2.5 times the tokens, so we did not run it on the remaining pairs.
For Sonnet 5.5, xhigh effort in both phases on the random sample of 49 pairs confirms a candidate on a ground-truth statement
in 36 functions, against 22 with medium effort and 36 for Opus 5.5 with medium effort,
at three times the cost of Opus 5.5 and four times the runtime.
For Opus 5.5, xhigh effort raises the hit rate from 66\% to 73\% (131 to 146 functions) and
MCC from 0.529 to 0.552 at almost the same precision (0.502 and 0.511),
but it confirms candidates in slightly more patched functions (54\% vs.\ 52\%), at 3.7 times the
runtime and 3.5 times the cost.

\paragraph{Differences between CWE types}
Out-of-bounds accesses (75 pairs) reach hit rates of 33--35\% for Qwen 3.8,
67\% for DeepSeek V4.1, 56\% for Sonnet 5.5, and 69\% (medium) and 84\% (xhigh) for Opus 5.5.
Unlike in Phase~1, NULL pointer dereferences (14 pairs) are verified about as well or
better (21--29\% for Qwen 3.8, 57--86\% for the others).
For CWE-703 (16 pairs), Qwen 3.8 gets a TP in only one function,
DeepSeek V4.1 in 56\%, Sonnet 5.5 in 31\%, and Opus 5.5 in 44\% and 50\%.
Rare types remain difficult to detect, with hits in at most one of the five
resource management pairs and one of the six reachable assertions.
Access-control and authentication weaknesses (7 pairs) are an exception for
Opus 5.5, which gets a TP in six (medium) and five (xhigh) of them,
whereas Qwen 3.8 gets none, DeepSeek V4.1 two, and Sonnet 5.5 one.

\paragraph{Answer to RQ2}
Phase~2 of \apn{} confirms at least one ground-truth statement with an accepted explanation in 26--73\% of the vulnerable functions.
Opus 5.5 performs best with a hit rate of 66\% (medium) and 73\% (xhigh) and a recall of 0.566 and 0.605,
followed by DeepSeek V4.1 (57\%, 0.466), Sonnet 5.5 (43\%, 0.366), and Qwen 3.8 (26\%, 0.242).
Higher reasoning effort improves Opus 5.5's hit rate but barely changes Qwen 3.8's.
Over the vulnerable and patched functions, MCC is 0.26--0.43.
However, all models also confirm candidates in many patched functions (23\% for Qwen 3.8, 37--61\% for the others),
limiting the function-level classification derived from confirmed verdicts. RQ3 examines this limitation.

\subsection{Results and Answer to RQ3}
\label{sec:rq3}

\tabref{tab:rq3-phase3} shows the results of the classification step based on the verdicts from Phase~2
under the three evaluation criteria presented in \secref{sec:setup:metrics}.
Label only compares the predicted labels with the ground-truth labels, regardless of the verdicts behind them.
Right reason (RR) additionally requires a vulnerable function classified as vulnerable to have a Phase~2 TP.
For CORRECT, which is more lenient than RR, an accepted explanation suffices regardless of the line,
and a patched function classified as vulnerable counts as an FP only if one of its explanations points
at the fixed vulnerability.

\begin{table*}[t]
\centering
\caption{Classification results on the 200 pairs per model and reasoning effort.}
\label{tab:rq3-phase3}
\footnotesize
\setlength{\tabcolsep}{3pt}
\begin{tabular}{l | l | rrrrrrrr|rrrrr|rrr}
\hline
 & & \multicolumn{8}{c|}{Label only} & \multicolumn{5}{c|}{Right reason} & \multicolumn{3}{c}{CORRECT} \\
Model & Effort & P-C & P-V & P-B & P-R & P & R & F1 & MCC & P-C$_{RR}$ & P$_{RR}$ & R$_{RR}$ & F1$_{RR}$ & MCC$_{RR}$ & P-C$_{C}$ & F1$_{C}$ & MCC$_{C}$ \\
\hline
Qwen $3.8_{LC}$ & medium & 47 & 36 & 107 & 10 & \textbf{0.643} & 0.415 & 0.505 & 0.198 & 35 & 0.531 & 0.260 & 0.349 & 0.035 & 54 & 0.428 & 0.379 \\
\hline
DeepSeek $V4.1_{LC}$ & high & 59 & 108 & 20 & 13 & 0.580 & 0.835 & 0.684 & 0.256 & 42 & 0.483 & 0.565 & 0.521 & $-$0.041 & 115 & 0.746 & 0.608 \\
\hline
Sonnet $5.5_{CC}$ & medium & 52 & 65 & 74 & 9 & 0.613 & 0.585 & 0.598 & 0.215 & 44 & 0.537 & 0.430 & 0.478 & 0.061 & 85 & 0.616 & 0.472 \\
\hline
Opus $5.5_{CC}$ & medium & 78 & 99 & 18 & 5 & 0.630 & 0.885 & 0.736 & 0.399 & 65 & 0.557 & 0.655 & 0.602 & 0.137 & 135 & 0.820 & 0.680 \\
\cline{2-18}
 & xhigh & \textbf{84} & 103 & 8 & 5 & 0.634 & \textbf{0.935} & \textbf{0.756} & \textbf{0.449} & \textbf{73} & \textbf{0.575} & \textbf{0.730} & \textbf{0.643} & \textbf{0.197} & \textbf{151} & \textbf{0.868} & \textbf{0.747} \\
\hline
\end{tabular}
\end{table*}

\paragraph{Label only pair-wise classification}
Opus 5.5 classifies the most pairs correctly, P-C of 84 (42\%) with xhigh and 78 (39\%) with medium effort,
with an MCC of 0.449 and 0.399. DeepSeek V4.1 classifies 59 pairs (29.5\%) correctly, Sonnet 5.5 52 (26\%),
and Qwen 3.8 47 (23.5\%).
The models fail in opposite ways. DeepSeek V4.1, Sonnet 5.5, and Opus 5.5 classify both functions as vulnerable in most of
their incorrect pairs (P-V of 65--108 pairs), whereas Qwen 3.8 classifies both as benign (P-B of 107 pairs).
Reversed pairs are rare (5--13 pairs).
Opus 5.5 classifies clearly more pairs correctly than DeepSeek V4.1 (78 and 84 vs.\ 59 pairs),
whereas DeepSeek V4.1, Sonnet 5.5, and Qwen 3.8 lie close together (47--59 pairs), and xhigh effort adds only
six correct pairs for Opus 5.5.
%Counting uncertain verdicts as positive generally lowers the number of correct pairs, because more patched than vulnerable functions change to a vulnerable classification.
%The exception is Qwen3.8, for which correct pairs increase from 47 to 55.

\paragraph{Correct for the right reason (RR)}
Correct labels do not always rest on correctly identified vulnerabilities, as also found by
Li \etal~\cite{li2025EverythingYouWantedKnow} and Ahmed \etal~\cite{ahmed2025SecVulEvalBenchmarkingLLMs}.
For Opus 5.5, 73 of the 84 (xhigh) and 65 of 78 (medium) pairs are also correct
for the right reason, for Sonnet 5.5 44 of 52 pairs, for Qwen 3.8 35 of 47 pairs,
and for DeepSeek V4.1 only 42 of 59 pairs.
Under the right-reason criterion, the lead of Opus 5.5 over DeepSeek V4.1 grows
(73 vs.\ 42 pairs with xhigh effort), and the lead of DeepSeek V4.1
over Sonnet 5.5 disappears (42 vs.\ 44 pairs). Recall drops from 0.415--0.935 to 0.260--0.730
and precision from 0.580--0.643 to 0.483--0.575,
since every patched function classified as vulnerable remains an FP.
MCC falls to 0.197 (xhigh) and 0.137 (medium) for Opus 5.5 and to around zero ($-$0.041 to 0.061) for the other configurations,
with Opus 5.5 retaining the strongest positive correlation with the labels.
Under the lenient CORRECT criterion,
DeepSeek V4.1, Sonnet 5.5, and Opus 5.5 classify 85--151 pairs correctly (F1-score 0.616--0.868). This criterion, however,
ignores most of the confirmed verdicts on patched functions that our statement-level metrics count as FPs.

\begin{table}[t]
\centering
\caption{Comparison with previous approaches on the PrimeVul Paired test set (previous approaches: all 435 pairs, \apn{}: our 200 pairs). P-C in \%.}
\label{tab:rq3-sota}
\footnotesize
\setlength{\tabcolsep}{1.7pt}
\begin{tabular}{l l | r r r | r r r}
\hline
 & & \multicolumn{3}{c|}{Label only} & \multicolumn{3}{c}{CORRECT} \\
Approach & Model & P-C & F1 & MCC & P-C & F1 & MCC \\
\hline
VulTrial & GPT-4o & 18.6 & 0.563$^*$ & 0.072$^*$ & -- & -- & -- \\
VulTriage & GPT-4o & 33.8 & 0.691 & 0.290$^*$ & -- & -- & -- \\
VulInstruct & DeepSeek-V3 & -- & -- & -- & 17.2 & 0.450 & 0.083$^*$ \\
\hline
\apn{} & Qwen 3.8 (med.) & 23.5 & 0.505 & 0.198 & 27.0 & 0.428 & 0.379 \\
 & DeepSeek V4.1 (high) & 29.5 & 0.684 & 0.256 & 57.5 & 0.746 & 0.608 \\
 & Sonnet 5.5 (med.) & 26.0 & 0.598 & 0.215 & 42.5 & 0.616 & 0.472 \\
 & Opus 5.5 (med.) & 39.0 & 0.736 & 0.399 & 67.5 & 0.820 & 0.680 \\
 & Opus 5.5 (xhigh) & \textbf{42.0} & \textbf{0.756} & \textbf{0.449} & \textbf{75.5} & \textbf{0.868} & \textbf{0.747} \\
\hline
\end{tabular}
\end{table}

\paragraph{Comparison with previous approaches}
\tabref{tab:rq3-sota} compares \apn{} with the three state-of-the-art approaches VulTrial~\cite{widyasari2025letTrialBegin},
VulTriage~\cite{tang2026VulTriageTriplePathContext}, and VulInstruct~\cite{zhu2025VulInstructTeachingLLMs}
based on the results reported in the corresponding papers. All three approaches were evaluated on the PrimeVul Paired
test set which allows a comparison with our approach. $^*$ denotes values computed by us from the reported
pair counts (VulTrial, VulTriage) or accuracy, precision, and recall (VulInstruct).
With both reasoning efforts, Opus 5.5 exceeds the reported P-C, F1-score, and MCC of previous approaches.
With xhigh effort, it achieves 42\% correct pairs, an F1-score of 0.756, and an MCC of 0.449,
compared with 33.8\%, 0.691, and 0.290 for VulTriage.
All configurations of \apn{}, including the open-weight models, classify more pairs
correctly than VulTrial (23.5--42\% vs.\ 18.6\%) with a higher MCC (0.198--0.449 vs.\ 0.072).
VulTriage classifies both functions as vulnerable in 47.6\% of pairs,
while DeepSeek V4.1 does so in 54\%. DeepSeek V4.1 reaches almost the same F1-score (0.684), despite an MCC of only 0.256.
Under the CORRECT criterion, all configurations outperform VulInstruct in P-C (27--75.5\% vs.\ 17.2\%)
and MCC (0.379--0.747 vs.\ 0.083), and all except Qwen 3.8 also in F1-score (0.616--0.868 vs.\ 0.450).
%Under our stricter right-reason criterion, 36.5\% (Opus 5.5) and 21\% (DeepSeek) of the
%pairs remain correct (\tabref{tab:rq3-phase3}).
%These numbers are not directly comparable because the
%models differ and our evaluation covers 200 of the 435 pairs (\secref{sec:discussion}).
%We do not compare with Phoenix~\cite{wang2026securityRelativeTrainingFree}, because it derives
%its specifications from both functions of each test pair (\secref{sec:related}).

\paragraph{Answer to RQ3}
\apn{} classifies 23.5--42\% of the pairs correctly, with Opus 5.5 performing best and the open-weight
models reaching 29.5\% (DeepSeek V4.1) and 23.5\% (Qwen 3.8). Of these correct pairs, 35--73 are also correct
for the right reason. Opus 5.5 achieves the most such pairs 73 (36.5\%) of all pairs.
The larger models DeepSeek V4.1, Sonnet 5.5, and Opus 5.5 mainly err by classifying patched functions as
vulnerable, while Qwen 3.8 often classifies both functions as benign.
Opus 5.5 exceeds the reported P-C, F1-score, and MCC of three recent approaches, and all configurations
exceed VulTrial and VulInstruct in P-C and MCC.

\section{Discussion}
\label{sec:discussion}

\subsection{Implications}
In the following, we discuss potential implications of our results on developers and the research on vulnerability detection.

\paragraph{For developers}
\apn{} can support security code reviews and the triage of software vulnerabilities.
With Opus 5.5 (xhigh), Phase~2 confirms 450 statements in the vulnerable functions (TP + FP in \tabref{tab:rq2-phase2}), 2.3 per function,
but also 281 in the patched functions, 1.4 per function.
Of these 731 confirmed statements, 230 are TPs, meaning about a third of them point a developer to the vulnerability (precision 0.315).
It misses 150 of the 380 ground-truth statements (FN, recall 0.605) and finds none of them in 54 of the 200 vulnerable functions (hit rate 73\%).
Each verdict output by \apn{} names the statement and explains why it is exploitable, so a developer can check the statement instead of searching the
function for the reason of a label. SAST tools leave developers with longer lists. In the comparison of Hassler \etal,
only 0.24\% of the reports of the best analyzer correspond to a known bug~\cite{hassler2025comparativeStudyFuzzers},
against 31.5\% of the statements confirmed by Opus 5.5 (xhigh) over the vulnerable and patched functions.
The settings differ, and our balanced sample contains far more vulnerable functions than real code. With 3\% vulnerable functions
as in PrimeVul~\cite{ding2025vulnerabilityDetectionCode}, the same rates would give a precision of about 2.4\%, which is still ten times higher than 0.24\%.
Even when running \apn{} with Qwen 3.8, the precision would be about 2.6\%, because it confirms fewer statements in patched functions (104 vs.\ 281).
However, Qwen 3.8 also finds fewer of the vulnerable statements (recall 0.242 vs.\ 0.605).

If the task of a developer is to detect vulnerable functions, \apn{} with Opus 5.5 (xhigh) flags 94\% of the vulnerable but also 54\% of the patched functions
(precision 0.634 in \tabref{tab:rq3-phase3}).
With 3\% vulnerable functions in a project, this would give a precision of about 5\%, \ie a developer would inspect about 20 flagged functions per vulnerable one.
In return, \apn{} misses only 13 of the 200 vulnerable functions (recall 0.935).
Under the right-reason criterion, recall drops to 0.730. In 146 of the 187 flagged vulnerable functions (hit rate 73\% in \tabref{tab:rq2-phase2}),
a confirmed statement points the developer to the vulnerability, and in the remaining 41, the developer still has to search the function for it.
With 3\% vulnerable functions, the right-reason precision of 0.575 would give about 4\%, \ie about 25 flagged functions per such vulnerable function.

\paragraph{For researchers}
Our results indicate that detecting vulnerabilities at the statement level is feasible with current LLMs and makes detection checkable.
With Opus 5.5 (xhigh), \apn{} confirms a ground-truth statement in 74\% of the vulnerable functions, and across all models, the judge accepts 94--99\% of the
confirmed verdicts on ground-truth statements, so a hit almost always comes with the right reason.
Verifying candidates in a second phase raises precision by a factor of 1.7--2.9 at a cost of 0.05--0.21 recall,
and the statement-level verdicts also yield the best pair-wise classification (42\% correct pairs, MCC 0.449).
Two limitations remain. First, all models also confirm statements in 23--61\% of the patched functions,
\ie they often do not recognize that the patch removed the vulnerability.
This is in line with Ding \etal~\cite{ding2025vulnerabilityDetectionCode} and Steenhoek \etal~\cite{steenhoek2024errMachineVulnerability},
who found that LLMs classifying functions directly hardly distinguish vulnerable from patched code.
Our results show that this limitation persists even when every verdict must be backed by evidence collected with tools.
Second, the model matters more than the amount of tool use. With medium effort and only 2--5 tool calls per function in Phases~1 and~2,
Opus 5.5 outperforms DeepSeek V4.1 (high), which makes 25--33 tool calls.

\subsection{Threats to Validity}
We identify several threats to construct, internal, and external validity of our results.

\paragraph{Construct validity}
Our statement-level metrics are strict in two ways and may therefore underestimate the performance.
First, a confirmed verdict counts as a TP only if it lies on a ground-truth statement, although the judge accepts the explanations of
another 30--84 confirmed verdicts per configuration that name the root cause but point at a different line.
Second, every confirmed verdict on a patched function counts as an FP, although a patched function may still contain other vulnerabilities.
The lenient CORRECT criterion in RQ3 relaxes both rules.
The LLM judge (Sonnet 5.5, high effort) belongs to the same model family as two of the evaluated models and might prefer their explanations.
To check this, we also judged the verdicts of Sonnet 5.5 and Opus 5.5 with DeepSeek V4.1. The two judges agree on 96--98\% of the confirmed
verdicts on vulnerable functions (Cohen's $\kappa$ 0.93--0.95), and Sonnet 5.5 accepts only 2--6 more verdicts per configuration,
which changes the TPs by at most four.

\paragraph{Internal validity}
Using Opus 5.5 to create the ground truth may favor statements that this model considers exploitable.
To mitigate this threat, the ground-truth tool received the CVE description, the commit message, and the diff, which \apn{} never sees.
Furthermore, two authors verified every function and resolved all disagreements by consensus.
Since most CVEs date from 2021 and 2022, before the training cut-off of all models, the models may know the vulnerabilities.
However, the agents never see the CVE, the commit, or the version history, and the many confirmed verdicts on patched functions indicate that
the models do not simply recall the fix.
We ran each configuration once, and the APIs of Anthropic and DeepSeek offer no seed, limiting reproducibility.
Differences of a few percentage points between configurations may therefore lie within this run-to-run variation.
Finally, a difference between models may partly stem from our two implementations of the agents.
However, both versions of \apn{} receive the same instructions only adapted to the available tools and equivalent filesystem and
text search tools. To identify potential differences, we ran the LangChain Phase~1 agent with Sonnet 5.5 which reached a
similar Phase~1 hit rate on 97 pairs compared to the implementation with the Claude Code Agent SDK (67\% vs.\ 65\%).

\paragraph{External validity}
We evaluated 200 randomly selected pairs of the PrimeVul Paired test set with C/C++ functions only.
Therefore, our results are valid for these 200 pairs, while previous approaches used all 435 pairs, which might bias the comparison with these approaches.
The 200 pairs cover 104 of the 123 projects and 45 of the 57 CWEs from the original PrimeVul Paired test set.
While we consider this a representative sample of real-world functions, our results remain limited to C/C++ projects.
Furthermore, we performed the experiments with four models, and the results might differ for other models.
We mitigated this threat by selecting two recent and highly rated open-weight models of different sizes
from two providers and included two recent frontier models from Anthropic.
Since new models are released frequently, we report their versions
and provide all prompts, tools, and results in our replication package.

\section{Conclusion}
\label{sec:conclusion}

We presented \apn{}, an agentic approach that detects exploitable statements in C/C++ functions in two phases. An agent first identifies candidate
statements, and a second agent analyzes each of them in detail and either confirms or rules it out with evidence, or marks it uncertain.
On a random sample of 200 pairs of the PrimeVul Paired test set with a manually verified ground truth,
Phase~2 confirms a ground-truth statement with an accepted explanation in up to 73\% of the vulnerable functions.
Because each confirmed verdict can be checked at its line, a function flagged as vulnerable becomes verifiable for developers.

The classification derived from these verdicts outperforms existing approaches, with up to 42\% correct pairs
and an MCC of 0.449 compared to 33.8\% and 0.290 for VulTriage~\cite{tang2026VulTriageTriplePathContext}, the best of them.
However, separating vulnerable from patched code remains challenging. Even the best-performing model, Opus 5.5,
confirms at least one candidate in 54\% of the patched functions, resulting in a statement-level precision of 0.315,
recall of 0.605, and F1-score of 0.414 over both vulnerable and patched functions.

In future work, we plan to first extend the evaluation of our approach to non-paired vulnerable and benign functions
from the PrimeVul dataset. Next, motivated by VulTriage, we plan to improve our approach by considering
information on vulnerabilities from the CWE database. Furthermore, we plan to investigate the benefits of equipping
our agents with static code analysis tools, such as codebadger~\cite{lekssays2026bridging}.

\section*{Use of Generative AI}
We used Anthropic Claude and OpenAI GPT models to assist with writing, coding, and literature search.
The authors reviewed and refined the generated material and take full responsibility for the final manuscript

%\section*{Acknowledgment}
%
%The preferred spelling of the word ``acknowledgment'' in America is without
%an ``e'' after the ``g''. Avoid the stilted expression ``one of us (R. B.
%G.) thanks $\ldots$''. Instead, try ``R. B. G. thanks$\ldots$''. Put sponsor
%acknowledgments in the unnumbered footnote on the first page.

%\section*{References}

\printbibliography

%\vspace{12pt}
%\color{red}
%IEEE conference templates contain guidance text for composing and formatting conference papers. Please ensure that all template text is removed from your conference paper prior to submission to the conference. Failure to remove the template text from your paper may result in your paper not being published.

\end{document}